\documentclass{elsart}

\usepackage{graphics}
\usepackage{amsmath}
\usepackage{times}
\usepackage{xspace}
\usepackage{array}
\usepackage{epsfig}

\newcommand{\dsubs}{\ensuremath{\mbox{D}^+_\mathrm{s}}\xspace}

\newcommand{\kzero}{\ensuremath{\mbox{K}^{0}}\xspace}

\newcommand{\mdzero}{\ensuremath{\mathrm{D}^{0}}\xspace}
\newcommand{\dzero}{\ensuremath{\mbox{D}^{0}}\xspace}
\newcommand{\dplus}{\ensuremath{\mbox{D}^{+}}\xspace}

\newcommand{\lbzero}{\ensuremath{\mathrm{\Lambda}^0}\xspace}
\newcommand{\lbcpl}{\ensuremath{\mathrm{\Lambda}^+_\mathrm{c}}\xspace}

\newcommand{\numu}{\ensuremath{\nu_\mu}\xspace}
\newcommand{\numubar}{\ensuremath{\overline{\nu}_\mu}\xspace}
\newcommand{\nutau}{\ensuremath{\nu_\tau}\xspace}
\newcommand{\nue}{\ensuremath{\nu_\mathrm{e}}\xspace}
\newcommand{\nubare}{\ensuremath{\overline{\nu}_e}\xspace}

\newcommand{\kg}{\ensuremath{\mbox{kg}}\xspace}
\newcommand{\GeV}{\ensuremath{\mbox{GeV}}\xspace}
\newcommand{\GeVc}{\ensuremath{\mbox{GeV}/\mbox{c}}\xspace}

\newcommand{\eVeVcccc}{\ensuremath{\mbox{eV}^2/\mbox{c}^4}\xspace}

\newcommand{\MeVc}{\ensuremath{\mbox{MeV}/\mbox{c}}\xspace}

\newcommand{\cm}{\ensuremath{\mbox{cm}}\xspace}
\newcommand{\mm}{\ensuremath{\mbox{mm}}\xspace}
\newcommand{\mmsq}{\ensuremath{\mbox{mm}^2}\xspace}
\newcommand{\micron}{\ensuremath{\mu \mbox{m}}\xspace}

\newcommand{\mrad}{\ensuremath{\mbox{mrad}}\xspace}

\newcommand{\veeTwo}{\ensuremath{\mathrm{V2}}\xspace}

\begin{document}

\begin{frontmatter}

\title{Final results on $\nu_\mu \rightarrow \nu_\tau$ oscillation from the CHORUS experiment}


\author{E.~Eskut, A.~Kayis-Topaksu, G.~\"{O}neng\"ut}
\address{\c{C}ukurova University, Adana, Turkey}

\author{M.G.~van Beuzekom, R.~van Dantzig, M.~de Jong, J.~Konijn}
\author{O.~Melzer, R.G.C.~Oldeman$^1$, E.~Pesen, C.A.F.J.~van der Poel}
\author{J.L.~Visschers}
\address{NIKHEF, Amsterdam, The Netherlands}

\author{M.~G\"uler, U.~K\"ose, M.~Serin-Zeyrek, R.~Sever, P.~Tolun, M.T.~Zeyrek}
\address{METU, Ankara, Turkey}

\author{N.~Armenise$^2$, F.~Cassol, M.G.~Catanesi, M.~De Serio, M.T.~Muciaccia}
\author{E.~Radicioni, P.~Righini, S.~Simone, L.~Vivolo}
\address{Universit\`a di Bari and INFN, Bari, Italy}

\author{A.~B\"ulte, K.~Winter}
\address{ Humboldt Universit\"at, Berlin, Germany}

\author{M.~Vander Donckt$^3$, B.~Van de Vyver, P.~Vilain, G.~Wilquet}
\address{Inter-University Institute for High Energies (ULB-VUB) Brussels, Belgium}

\author{B.~Saitta}
\address{Universit\`a di Cagliari and INFN, Cagliari, Italy}

\author{E.~Di Capua, C.~Luppi}
\address{Universit\`a di Ferrara and INFN, Ferrara, Italy}

\author{Y.~Ishii, M.~Kazuno, S.~Ogawa, H.~Shibuya}
\address{Toho University,  Funabashi, Japan}

\author{J.~Brunner, M.~Chizhov, D.~Cussans, M.~Doucet, J.P.~Fabre, W.~Flegel}
\author{I.R.~Hristova$^4$,  T.~Kawamura, D.~Kolev$^5$, M.~Litmaath, H.~Meinhard}
\author{E.~Niu, H.~\O ver\.{a}s, J.~Panman, I.M.~Papadopoulos, S.~Ricciardi$^6$}
\author{A.~Rozanov$^7$, D.~Saltzberg$^{8}$, R.~Tsenov$^5$, J.W.E.~Uiterwijk}
\author{C.~Weinheimer, H.~Wong, P.~Zucchelli}
\address{CERN, Geneva, Switzerland}

\author{J.~Goldberg}
\address{Technion, Haifa, Israel}

\author{M.~Chikawa}
\address{Kinki University, Higashiosaka, Japan}

\author{E.~Arik, A.A.~Mailov}
\address{Bogazici University, Istanbul, Turkey}

\author{J.S.~Song, C.S.~Yoon}
\address{Gyeongsang National University,  Jinju, Korea}

\author{K.~Kodama, N.~Ushida}
\address{Aichi University of Education, Kariya, Japan}

\author{S.~Aoki, T.~Hara}
\address{Kobe University,  Kobe, Japan}

\author{G.~Brooijmans$^{9}$, T.~Delbar,  D.~Favart, G.~Gr\'egoire, J.~H\'erin, S.~ Kalinin$^{10}$}
\author{I.~ Makhlioueva}
\address{Universit\'e Catholique de Louvain, Louvain-la-Neuve, Belgium} 

\author{A.~Artamonov, P.~Gorbunov, V.~Khovansky, V.~Shamanov, I.~Tsukerman}
\address{Institute for Theoretical and Experimental Physics, Moscow, Russian Federation}

\author{D.~Bonek\"amper, N.~Bruski, D.~Frekers, D.~Rondeshagen, T.~Wolff}
\address{Westf\"alische Wilhelms-Universit\"at, M\"unster, Germany}

\author{K.~Hoshino, J.~Kawada, M.~Komatsu, Y.~Kotaka, T.~Kozaki}
\author{M.~Miyanishi, M.~Nakamura, T.~Nakano, K.~Narita, K.~Niu, K.~Niwa}
\author{N.~Nonaka, Y.~Obayashi, O.~Sato, T.~Toshito}
\address{Nagoya University, Nagoya, Japan}

\author{S.~Buontempo, A.G.~Cocco, N.~D'Ambrosio$^{11}$, G.~De Lellis}
\author{G.~ De Rosa, F.~Di Capua, A.~Ereditato$^{12}$, G.~Fiorillo, A.~Marotta}
\author{M.~Messina$^{12}$, P.~ Migliozzi, V.~Palladino, L.~Scotto Lavina, P.~Strolin}
\author{V.~Tioukov}
\address{Universit\`a Federico II and INFN, Naples, Italy}

\author{K.~Nakamura, T.~Okusawa}
\address{Osaka City University, Osaka, Japan}

\author{A.~Capone, D.~De Pedis, S.~Di Liberto, U.~Dore, P.F.~Loverre}
\author{L.~Ludovici, A.~Maslennikov, M.A.~Mazzoni, G.~Piredda, G.~Rosa}
\author{R.~Santacesaria, A.~Satta$^{13}$, F.R.~Spada}
\address{Universit\`a La Sapienza and INFN, Rome, Italy}

\author{E.~Barbuto, C.~Bozza, G.~Grella, G.~Romano, C.~Sirignano}
\author{S.~Sorrentino}
\address{ Universit\`a di Salerno and INFN, Salerno, Italy}

\author{Y.~Sato, I.~Tezuka}
\address{Utsunomiya University,  Utsunomiya, Japan}

\thanks[1]{Now at Universit\`a di Cagliari, Cagliari, Italy.}
\thanks[2]{Deceased.}
\thanks[3]{Now at CERN, Switzerland.}
\thanks[4]{Now at DESY, Hamburg.}
\thanks[5]{On leave of absence and at St. Kliment Ohridski University of Sofia, Bulgaria.}
\thanks[6]{Now at Rutherford Appleton Laboratory, Oxon, United Kingdom.}
\thanks[7]{Now at CPPM CNRS-IN2P3, Marseille, France.}
\thanks[8]{Now at University of California, Los Angeles, USA.}
\thanks[9]{Now at Columbia University, New York, USA.}
\thanks[10]{Now at Rheinisch-Westfaelische Technische Hochschule, Aachen, Germany.}
\thanks[11]{Now at Laboratorio Nazionale del Gran Sasso, Assergi, Italy.}
\thanks[12]{Now at Bern University, Bern, Switzerland.}
\thanks[13]{Now at Universit\`a di Roma Tor Vergata, Rome, Italy.}

\begin{abstract}
The final oscillation analysis of the complete set of data collected by CHORUS in the years 1994--1997 is presented.
Reconstruction algorithms of data extracted by electronic detectors were improved and the data recorded in the 
emulsion target were analysed by new automated scanning systems, allowing the use of a new method for event
reconstruction in emulsion. CHORUS has applied these new techniques to the sample of 1996--1997 events for which 
no muons were observed in the electronic detectors. 
Combining the new sample with the data analysed in previous papers, the overall sensitivity of the 
experiment to the $\nu_\tau$ appearance is thus improved.
In a two-neutrino mixing scheme, a 90\%~C.L.\ upper limit of $\sin^2 2\theta_{\mu\tau} < 4.4 \times 10^{-4}$ 
is set for large $\Delta m^2$, improving by a factor 1.5 the previously published CHORUS result.
\end{abstract}

\end{frontmatter}

\section{Introduction}

The CHORUS experiment was designed to search for \numu\ $\rightarrow$ \nutau\ oscillations
through the observation of charged-current interactions \nutau\ $N \rightarrow \tau^- X$
followed by the decay of the $\tau$ lepton, directly observed in a nuclear emulsion target.
The experiment aimed at achieving maximum sensitivity on the effective mixing angle for values of 
the mass parameter $\Delta m^2$ larger than $10$~\eVeVcccc.
This particular choice was based on the hypothesis that the neutrino mass could contribute 
to the solution of the Dark Matter puzzle \cite{harari,ellis}.
A short-baseline experiment in the CERN SPS Wide Band Neutrino Beam \cite{sps} was well suited
for this search. Another experiment, NOMAD, used the same beam and searched for $\nu_\tau$ 
appearance using a purely electronic technique \cite{nomad}.

The CHORUS experiment took data from 1994 to 1997.
A first phase of data analysis (`Phase I') was performed and no evidence for oscillations 
was found \cite{osc1,osc2,osc3}. 
Owing to several improvements in automated emulsion scanning and in the event recontruction,
it was considered worthwhile to perform a new and more complete analysis (`Phase II') 
of the data collected in the 1996--1997 period.

In this paper, we report the analysis performed on the events
for which no muons were observed in the electronic detector (\emph{$0\mu$ events}).
In particular, a dedicated search for decays into three charged hadrons was also
performed.
Combining this new sample with the data samples analysed and published in previous papers, {\em i.e.}
with a muon observed in the electronic detector ($1\mu$) for the whole period of data taking
and a small $0\mu$ sample collected in 1994--1995, leads to an improvement of the 
CHORUS sensitivity to \numu\ $\rightarrow$ \nutau\ appearance by a factor 1.5 and to 
\nue\ $\rightarrow$ \nutau\ appearance by a factor 1.2.

Although it is now established that \numu\ $\rightarrow$ \nutau\ oscillations occur at 
$\Delta m^2 \sim 10^{-3}$, this paper intends to show the capabilities 
of an hybrid emulsion experiment and that the goal for which CHORUS was designed has been reached.


\section{The experimental setup}

The CERN SPS Wide Band Neutrino Beam consists essentially of
\numu\, with a contamination of about 5.1\% \numubar, 0.8\% of \nue, and 0.2\%
of \nubare, while the flux of \nutau\ is
negligible~\cite {bart}. The average neutrino energy 
is 26~\GeV, well above the $\tau$ production threshold.

The CHORUS detector is a hybrid setup which combines a nuclear emulsion target
with various electronic detectors such as trigger hodoscopes, a scintillating
fibre tracker system, a hadron spectrometer, electromagnetic and hadronic
calorimeters, and a muon spectrometer~\cite{chorusdet}.  

Thanks to its micrometric granularity, nuclear emulsion allows a precise three-dimensional 
reconstruction of the neutrino interaction vertex as well as of the decay vertices of associated 
short-lived particles.
It is a powerful detector technique to directly observe $\tau$ decays which, 
in this experiment, occur on average at a distance of $1.7$ mm from the interaction vertex.

The scintillating fibre tracker system consists of two components: the target tracker
(TT), which is the major tool for locating in the emulsion the region where a neutrino 
interaction has occurred, and the diamond tracker (DT), a set of trackers associated with 
the hadron spectrometer.

The hadron spectrometer measures the charge and momentum of charged particles
using an air-core magnet. The calorimeter is used to determine the 
energy of electromagnetic and hadronic showers. The magnetized iron muon spectrometer determines
the charge and momentum of muons.

The emulsion target has an overall mass of 770~\kg and is segmented into four stacks. 
Each stack consists of eight modules with 36 plates of size 36~\cm$\times$~71~\cm. Each
plate has a 90~\micron plastic base coated on both sides with a 350~\micron 
emulsion layer. Each stack is followed by three interface emulsion
sheets having a 100~\micron emulsion layer on both sides of a 800~\micron
thick plastic base, and by a set of TT planes. 
The TT predicts particle trajectories on the interface emulsion sheets 
with a precision of about 150~\micron in position and 2~\mrad in track angle.

The emulsion scanning is performed by computer-controlled, fully
automated microscope stages equipped with a CCD camera and a read-out
system called `track selector'~\cite{aoki,TrackSelector}. 
The last generation of automated microscopes used in CHORUS experiment
is the UTS (`Ultra Track Selector')~\cite{UTS}.
In order to recognize track segments in the emulsion, a series 
of tomographic images is taken by focusing at different depths in the 
emulsion layer. 
The digitized images at different levels are shifted according to the 
predicted track angle and then added.  
The presence of aligned grains forming a track is detected as a local
peak in the grey-level of the summed image. 
The track-finding efficiency is higher than 98\% for 
track slopes less than 400~\mrad~\cite{TrackSelector}.


\section{The event reconstruction}
\label{sec:emulsion}

The event reconstruction algorithms have been improved significantly with 
respect to the Phase I analysis described in detail in Ref.~\cite{osc3}.
Here, we only summarize the essential points.

The event reconstruction starts with the pattern recognition in the electronic detectors.
Tracks are found in the TT, in the calorimeter and in the muon spectrometer. 
If a muon is detected in the downstream detectors 
the event is classified as $1\mu$, otherwise it is called $0\mu$.
Vertices are reconstructed by the electronic detectors using the points 
of closest approach of the tracks in the TT immediately downstream from the target.
The main vertex is the most upstream one and is selected if it contains either a muon
or a hadron surviving criteria which differed in the two phases
\begin{footnote}
{
 In Phase I, $0\mu$ events, at least one hadron with a momentum less
 than 20~\GeVc was required.  In Phase II the most isolated hadron was
 selected regardless of its charge and momentum.
}
\end {footnote}.
Such particles or the muon are used as so-called `scan-back' tracks. 
The impact points of the scan-back tracks are predicted on the most
downstream interface emulsion sheets, to initiate their follow back in emulsion.

As in previous analyses the event reconstruction in emulsion starts with a
procedure called `vertex location'. As a first step, all the tracks in the interface 
emulsion sheets within an area of 1~\mmsq centred on each scan-back track prediction 
are collected.

The correct association of emulsion tracks with TT tracks, both in position and direction, 
requires a precise alignment of the interface emulsions with respect to the fibre-trackers.
The alignment parameters are obtained by finding the best match between the
full set of TT predicted tracks and the full set of collected emulsion tracks.
Once an emulsion track is associated to a scan-back track, mainly on the basis of their 
common direction, it is followed upstream from one emulsion plate to the next within 
a greatly reduced scanning area as the resolution improves.
In the emulsion stacks, the scanning area reduces to a square of 50~\micron side.
The plate-to-plate alignment is first obtained by a coarse adjustment of reference marks, 
and, as the scanning procedure continues, is refined by aligning track
maps measured in subsequent plates. 
For this alignment we used tracks of
particles generated by neutrino interactions during the whole data taking, of cosmic rays 
and of muons coming from neighbouring beams.
The scan-back procedure stops when a searched track is not found in two consecutive plates. 
The most downstream one is defined as the "vertex plate".
This procedure for locating the vertex has an efficiency of 34\%
 (Table~\ref{table:0muReconstruction}). 

Once the vertex plate is found, a new scanning technique is applied on the 1996--1997 
data sample. Originally developed for the DONUT experiment~\cite{donut}, this technique, 
called `NetScan' \cite{nonaka}, is described in detail in Ref.~\cite{murat,luca}
together with the event location procedure. Its application to the $0\mu$ sample is 
also described in the following section
\begin{footnote}
{
The NetScan technique has been used also for the $1\mu$ data sample, mainly to search 
for charmed particles decays~\cite{d0,d0star}.
}
\end {footnote}.


\subsection{The NetScan technique}

Once the vertex plate is identified, the UTS performs a 
scan of the emulsion volume around the vertex position,
recording, for each event, all track segments within 400~\mrad with
respect to the orthogonal direction of the plates.
In each plate only the most upstream 100~\micron part is scanned.
The scanned volume is $1.5 \ \mm \times 1.5 \ \mm$ wide and 6.3~\mm long 
in the beam direction, corresponding to eight emulsion plates.
This volume contains the vertex plate itself, the plate immediately upstream, 
and the six plates downstream the vertex plate.
The plate upstream the vertex acts as a veto for passing through tracks.
The six plates downstream from the vertex act as decay volume and are
used to detect the tracks of the decay daughters.
The scan area is centred on the scan-back track stopping point. 

The number of track segments (coming from particles induced by neutrino 
interactions, cosmic ray particles and muons from neighbouring beams)
found in each plate depends on the position of the NetScan volume 
with respect to the beam centre and on average is $920$.

The task of the NetScan event reconstruction is to select
the segments belonging to the neutrino interaction
under study, out of this large number of background track segments. 
To this end, a first plate--to--plate alignment is performed by comparing the pattern of 
segments in a plate with the corresponding pattern in the next upstream plate.
Each segment found in one plate is
extrapolated to the next plate where a matching segment is looked for
within about 4~\micron in position (3$\sigma$ of alignment resolution) 
and 20 \mrad in angle. If none is found, the straight-line 
extrapolation is tried one plate further upstream.

A second and more accurate inter-plate alignment is performed using tracks 
passing through the entire volume after the connection of all matched
segments.
These tracks come from muons associated with the 
neutrino beam or with charged particle beams in the same experimental
area. 
After this fine alignment, the distribution of the residuals of the segment
positions with respect to the  
fitted track has a standard deviation width of about 0.45~\micron.

After rejection of isolated track segments, typically about 400 tracks 
remain in the volume. The majority of these are tracks 
(mainly Compton electrons and $\delta$-rays) with a 
momentum less than 100~\MeVc. These background tracks are rejected on the basis of 
the $\chi^2$ of a straight-line fit to the track segments.
The final step is the 
rejection of the tracks not originating from the scan volume. After this 
filtering, the average number of remaining tracks is about 40.
The number of connected tracks and the position and slope residuals provide a good 
measurement of the alignment quality. About 84\% of the events passes
the quality cuts (`NetScan acquisition accepted').

The reconstruction algorithm then tries to associate the tracks to vertices with 
a `pair-based' method: after selecting pairs of tracks having minimal 
distance less than 10~\micron, a clustering among them is performed and 
vertex points are defined.
After this clustering, a track is attached to a vertex if its distance from the 
vertex point (called hereafter `impact parameter') is less than 3~\micron. 
At the end of the procedure, one defines a primary vertex (and 
its associated tracks) and in case one or more secondary vertices to 
which `daughter tracks' are attached.

About 77\% of the events, where the NetScan acquisition was accepted,
have a vertex reconstructed in the NetScan volume. 
The loss is accounted for by the fact that a background track 
might have been selected for the scan-back procedure.
The numbers of events at the various stages of the procedure are given in 
Table~\ref{table:0muReconstruction}.

\begin{table}[hbpt]
  \begin{center}
    \caption{Results of the reconstruction of the $0\mu$ sample}
    \label{table:0muReconstruction}
    \begin{tabular}{cc} 
      {\bf Stage of reconstruction}      & {\bf number of events} \\ \hline
      Interface emulsion scanned         & 102544 \\ 
      Vertex plate found                 &  35039 \\
      NetScan acquisition accepted       &  29404 \\
      Vertex reconstructed               &  22661 \\ 
    \end{tabular}
  \end{center}
\end{table}


\subsection {The decay search}
\label {sec:decay}

\begin{figure}[ht]
  \begin{center}
      \resizebox{0.6\textwidth}{!}{ 
	\includegraphics{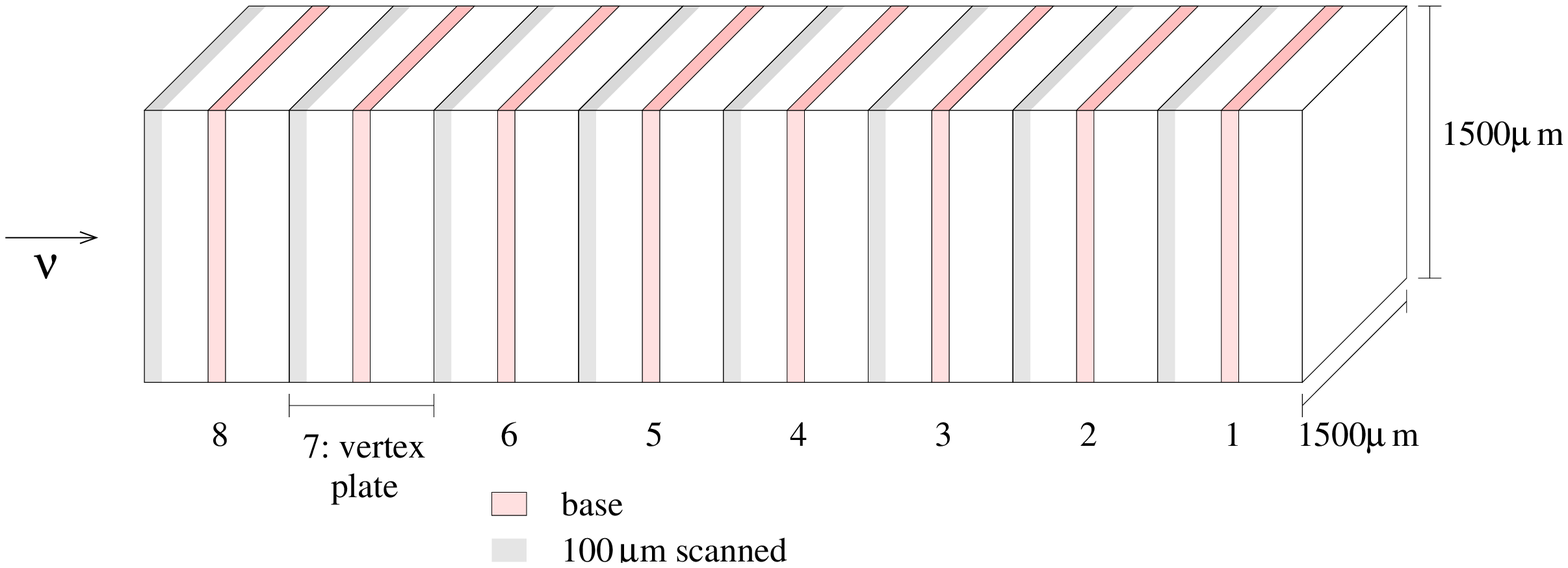}}
      \resizebox{0.5\textwidth}{!}{ 
	\includegraphics{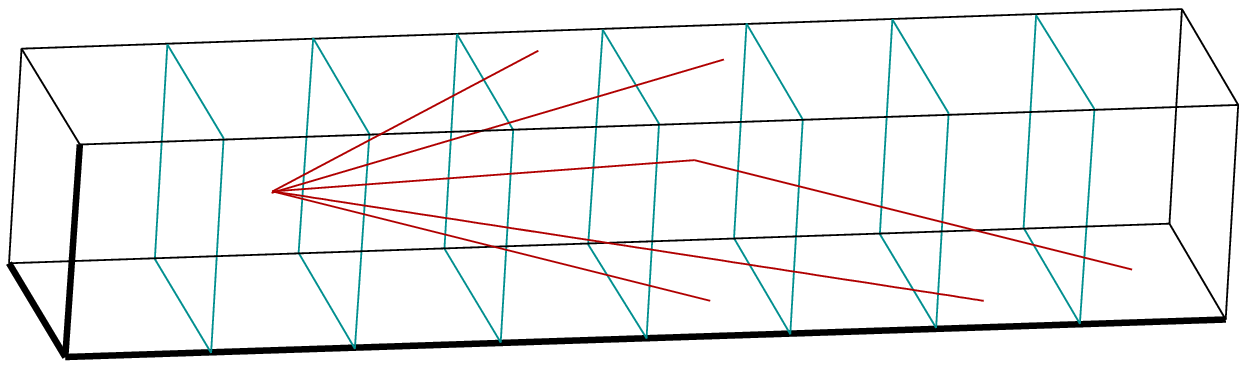}}
    \caption{The NetScan volume and a Monte Carlo event with one-prong ({\em i.e.} kink) $\tau$ decay.}
    \label{fig:netscanevent}
  \end{center}
\end{figure}

Once tracks and vertices are reconstructed in the NetScan volume, 
a search for decay topologies is performed.
In Fig.~\ref{fig:netscanevent} a sketch of a $\nu_\tau$ 
Monte Carlo event inside  the NetScan volume is shown: a $\nu_\tau$ (not drawn) 
interacts producing a $\tau$ lepton and other particles. 
After three plates, the $\tau$ lepton decays producing one charged particle 
(kink topology, {\em i.e.} an observed abrupt change of direction in the track).

In the decay search an emulsion track is called a `TT-confirmed track' if its trajectory 
matches that of a TT reconstructed track, or if it is connected to a sequence of TT hits. A reconstructed 
vertex is a `TT-confirmed vertex' if it contains at least one TT-confirmed track.

The decay search is done in two stages:

\begin{itemize}
\item[i)] an automated search on the whole sample;
\item [ii)] a computer-assisted visual inspection on the sample
selected in the previous stage.
\end{itemize}

After the automated search, 754 events are selected as belonging to one of three categories:

\begin{itemize}
\item{\emph{Short decays.}}
Events for which there is only one TT-confirmed vertex in the NetScan volume and at least
one isolated TT-confirmed track having an impact parameter with the primary vertex inside the range
$[7.2,100]$~\micron.
\item{\emph{Long decays.}}
Events with a one-prong decay topology (`kink') directly observed in the NetScan volume.
This occurs when the parent track $\tau$ is observed in at least one emulsion plate.
The daughter track must be TT-confirmed and its minimal distance to the parent track 
is required smaller than $6$~\micron. In case a primary vertex is reconstructed, 
it is required to be TT-confirmed and the parent track must have an impact parameter 
less than $3$~\micron with this vertex.
\item{\emph{Multiprong decays.}}
Events in which NetScan reconstructs at least two vertices, the secondary vertices being candidates 
for multiprong $\tau$ decays. The quality of the vertices is ranked according to the number 
of TT-confirmed tracks originating from them and it is used as a parameter for the selection of 
multiprong decay topology. The parent angle, in the case of a neutral particle decay 
deduced from the line connecting the primary and secondary vertices, should be within
400 \mrad from the beam direction. 

\end{itemize}

The computer-assisted visual inspection aims at clearly establishing the topology of the primary 
and secondary vertices. To avoid ambiguous topologies, the flight length of the parent track 
is required to be larger than 25~\micron. For vertices which are not both TT-confirmed, 
a more stringent cut, which is a function of the flight length of the assumed parent, is applied to 
remove the random association of two unrelated vertices.

An event is selected if one of the following topologies is detected:

\begin{itemize}
\item C1 : a track with a kink (interaction or decay) of at least 50 mrad
\item C3 : a track with a three-prong (interaction or decay) topology
\item V2 : a two-prong neutral interaction (or decay) is detected
\end{itemize}

The numbers of events observed in these topologies are shown in Table~\ref{table:SelectedEvents}.
Decays of the $\tau$ lepton are characterized by either one or three
charged particles in the final state, C1 or C3. The events with a V2 topology will be used for 
background calculations.

\begin{table}[htb]
  \caption{ $0\mu$ events selected after visual inspection.}
  \begin{displaymath}
    \begin{array}{cc}
     \textrm{\bf Topology} & \textrm{\bf number of events}\\ 
     \hline
     \textrm{C1} & 59  \\
     \textrm{C3} & 48  \\
      \textrm{V2} & 99  \\
    \end{array}
  \end{displaymath}
  \label{table:SelectedEvents}
\end{table}


\subsection{Background evaluation}

Physical processes that can mimic a $\tau$ decay are:
\begin{itemize}
\item decays of charmed charged particles like \dplus, \dsubs and \lbcpl, if the primary $\mu^-$ is not 
identified and the charge of the charmed particle is not determined;
\item white interactions, {\em i.e.} interactions of hadrons without any other visible activity at the 
vertex (recoil or Auger electrons);
\item $\Sigma$ decays which affect only the C1 topology.
\end{itemize}

The charmed particle contribution to the background has been evaluated from
production cross-section of \dplus, \dsubs and \lbcpl in neutrino interactions and
their decay branching fractions into one or three charged particles \cite{muonicbr}.
The rate of \dzero production in neutrino charged-current interactions 
has been measured in CHORUS~\cite{d0}. 
Using as the normalization the observed number of events with a V2 topology given 
in Table~\ref{table:SelectedEvents}, the number of one and three-prong decays from 
charm particles in the $0\mu$ sample is given by


$$ N_{Ci}=N_{\veeTwo}
\cdot \frac{\sigma_{\mathrm{C}}^{+}}{\sigma_{\mdzero}}
\cdot \frac{\epsilon_{\mathrm{C}i}}{\epsilon_{\veeTwo}}~,
$$

\noindent where $i=1$ or $3$, $\sigma_{\mathrm{C}}^{+}/\sigma_{\mdzero}=1.03\pm0.16$ is the ratio
of charged charm and \dzero cross-sections  \cite{muonicbr}
and $\epsilon _{\veeTwo}$, $\epsilon_{\mathrm{C1}}$ and $\epsilon_{\mathrm{C3}}$
are the efficiencies in detecting the specific topology, including the branching ratios,
computed by a Monte Carlo simulation: $\epsilon_{\mathrm{C1}}/\epsilon_{\veeTwo} = 0.28\pm0.08$, 
$\epsilon_{\mathrm{C3}}/\epsilon_{\veeTwo} = 0.45\pm0.09$.

The number of V2 events has been corrected to account for \kzero and \lbzero decays \cite{luca}.
This yields a prediction of $N_\mathrm{C1} = 27.2\pm 8.6$ and $N_\mathrm{C3} = 44 \pm 11$.

One-prong interactions without visible activity (`white kinks', WK) were
generated and processed through the full simulation chain assuming a hadron interaction
length $\lambda = 24$ m~\cite{satta}.
In the $1\mu$ sample, the ratio of three-prong to one-prong interactions, both with visible activity,
was measured to be $\sim 11\%$.  
This fraction was assumed to be the same also in
absence of recoil or Auger electrons~\cite{luca}.

The results are summarized in Table~\ref{table:DataAndBg}.

\begin{table}[hbt]
  \caption{ Expected background and observed data for C1 and C3 topologies in the $0\mu$ sample.}
  \begin{displaymath}
    \begin{array}{c|ccccc}
      \textrm{\bf Topology}&\textrm{\bf Charm}&\textrm{\bf
       White interactions}&\textrm{\bf Other decays}&\textrm{\bf Total
       BG}&\textrm{\bf Data}   \\
      \hline
      0\mu~\textrm{C1} &    27.2\pm8.6  &    24.9\pm2.7   &    1.11\pm0.34  & 53.2\pm9.0  &    59   \\
      0\mu~\textrm{C3} &    44\pm11     &  2.7\pm0.3   &        -        & 47\pm11  &    48   \\
    \end{array}
  \end{displaymath}
  \label{table:DataAndBg}
\end{table}


\subsection{Maximum number of detectable $\tau$}

We define $N_\tau^\mathrm{MAX}$ as the number of $\tau$ events that would be observed if their
detection efficiency and the oscillation probability would be equal to 1. It can be evaluated as:

$$ N_\tau^\mathrm{MAX} = \int \Phi_{\nu_{\mu}}(E_\nu) \cdot \sigma_{\nu_{\tau}}(E_\nu)~dE = 
\frac{N_\mathrm{CC}}{\epsilon_\mathrm{CC}}
\cdot \frac{\langle \sigma_{\nu_{\tau}} \rangle}{\langle \sigma_{\nu_{\mu}} \rangle}~,
$$ 

\noindent where $\Phi_{\nu_{\mu}}(E_\nu)$ is the neutrino flux, $N_\mathrm{CC}$ (93807) is the number 
of charged-current (CC) neutrino interaction events (used only for the flux determination) 
detected in the 1996--1997 run for which NetScan acquisition is completed, and  
$\epsilon_\mathrm{CC}$ their detection efficiency. 
$\langle \sigma_{\nu_{\tau}} \rangle / \langle \sigma_{\nu_{\mu}} \rangle$ is the ratio
between the average total $\nu_{\mu}$ and $\nu_{\tau}$ CC cross-sections.
Efficiencies have been calculated by Monte Carlo. We define
$\epsilon^\mathrm{loc}$ as the probability to locate a vertex and
$\epsilon^\mathrm{sel}$ as the reconstruction and selection efficiencies for the
various decay channels. For each topology, the number $N_\tau^\mathrm{max}$
of $\tau$ events that would be observed if the oscillation probability would be equal to 1 is:

$$ N_\tau^\mathrm{max} = N_\tau^\mathrm{MAX} \cdot \sum_m {BR_m \cdot \epsilon^\mathrm{loc}_{m} \cdot \epsilon^\mathrm{sel}_{m}}~,$$ 

\noindent where the index $m$ runs over the various $\tau$ decay modes contributing to that topology. 
Branching ratios, efficencies and $N_\tau^\mathrm{max}$ are given in Table~\ref{table:Ntaumax}. 
Note that muonic decays with a misidentified muon also contribute to the C1 topology.

\begin{table}[ht]
  \caption{ $\tau$ branching ratios, efficiencies and maximum detectable $\tau$ events for C1 and C3 topologies in the $0\mu$ sample.}
  \begin{displaymath}
    \begin{array}{c|l|ccc|r|r}
      \textrm{\bf Topology} & \textrm{\bf Channel} & BR~(\%) & \epsilon^\mathrm{loc}~(\%) &
       \epsilon^\mathrm{sel}~(\%) & \multicolumn{2}{c}{N_\tau^\mathrm{max}} \\ 
      \hline
              & \tau \rightarrow \mathrm{h}    & 49.5 & 32.6\pm0.5 & 23.7\pm0.6 & 7227 &      \\
      0\mu~\textrm{C1} & \tau \rightarrow \mathrm{e}    & 17.8 & 26.9\pm0.6 & 21.2\pm1.3 & 1922 & 9621 \\ 
              & \tau \rightarrow \mu           & 17.4 & ~~5.9\pm0.3 &   24\pm3   &  472 &      \\ 
      \hline
      0\mu~\textrm{C3} & \tau \rightarrow \mathrm{3h}   & 15.2 & 35.8\pm0.8 & 43.1\pm1.5 & 4443 & 4443 \\ 
    \end{array}
  \end{displaymath}
  \label{table:Ntaumax}
\end{table}


\subsection{Post-scan selection and final data sample}
\label{sec:post-scanning}

The signal to background ratio, hence the sensitivity to oscillation, can be improved if, 
in addition to the topology, other information is used, such as:
\begin{itemize}
\item the angle between the parent track and the estimated direction of the hadronic shower
\item the length of the parent track
\item the mean angle of the daughter tracks with respect to the parent track
\item when available, the momentum and charge of a daughter particle
\end{itemize}
The purpose of the `post-scan' analysis is to maximize the sensitivity to oscillation 
by an optimum combination of this information. The optimization was performed on Monte Carlo 
simulated samples of signal and background events~\cite{luca} and led to the subdivision 
of the samples into different subclasses, depending on the observed topology.

\subsubsection{The C1 topology}

For the C1 topology, the transverse momentum with respect to the parent direction 
is useful for the rejection of WK background. 

Two methods are available for momentum measurements:
either with the DT of the hadron spectrometer or from the multiple Coulomb scattering 
(MCS) in the emulsion. The DT also measures the particle charge. 
The MCS method is used if no DT measurement is available and if the number of emulsion plates 
for the MCS measurement is greater than 6.
The C1 sample has thus been divided into three subclasses: 

\begin{itemize}
\item momentum measurement from DT ($P^{\mathrm{DT}}$) which also yields a charge measurement of the sign
of the charge which is important for the charm background rejection
\item momentum measurement from MCS ($P^{\mathrm{MCS}}$)
\item no momentum measurement
\end{itemize}

\begin{figure} [htp]
  \centerline{\psfig{figure=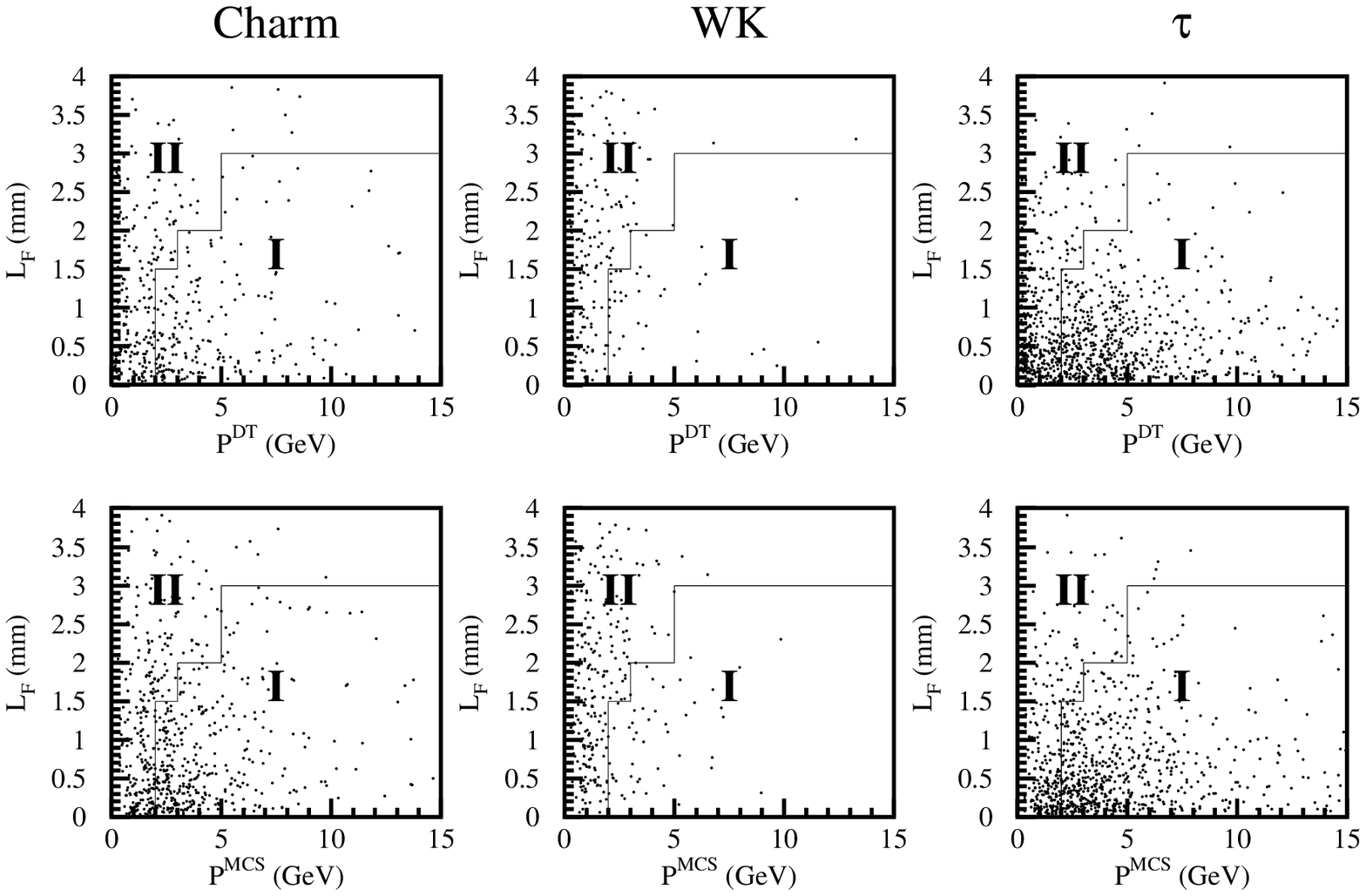,width=15.0cm}}
    \caption{The flight length $L_\mathrm{F}$ versus daughter particle momentum computed for charm (left), 
white kink (centre) and $\tau$ events (right). The momentum is evaluated by DT (top) or MCS (bottom). Only 
events in region I are selected.}
    \label{fig:fl_vs_p}
\end{figure}

The first category has the highest sensitivity to $\tau$ search and the strongest background 
rejection. The events in which the daughter track has a positive charge are rejected.
When the momentum is determined, a cut is applied on the transverse momentum ($P_{\mathrm{T}}>250$~\MeVc) 
of the daughter kink.
A momentum-dependent cut on the flight length ($L_\mathrm{F}$) is also applied. 
Fig.~\ref{fig:fl_vs_p} shows a comparison
between signal and background in the selected (I) and rejected (II) regions in
the $L_\mathrm{F}$ versus $P$ plane. 
All decay vertices beyond 3~\mm or with a parent momentum below 2~\GeVc
were rejected because white kink interactions are likely to populate these regions.
For a momentum between 2~\GeVc and 5~\GeVc step functions were used for
simplicity. 
The effects of this selection on
the background rejection and $\tau$ efficiency are shown in Table \ref{table:flvsp}.

\begin{table}[ht]
  \caption{ Effect of the selection in flight length-momentum plane on the signal and background, 
without considering any other cut.}
  \begin{displaymath}
    \begin{array}{c|c}
      \textrm{\bf Physical process} & \textrm{After rejection of events} \\ 
      \hline
      \textrm{WK}    & 13\% \\
      \textrm{charm} & 34\% \\
      \tau           & 52\% \\
    \end{array}
  \end{displaymath}
  \label{table:flvsp}
\end{table}

\begin{figure} [htp] 
  \centerline{\psfig{figure=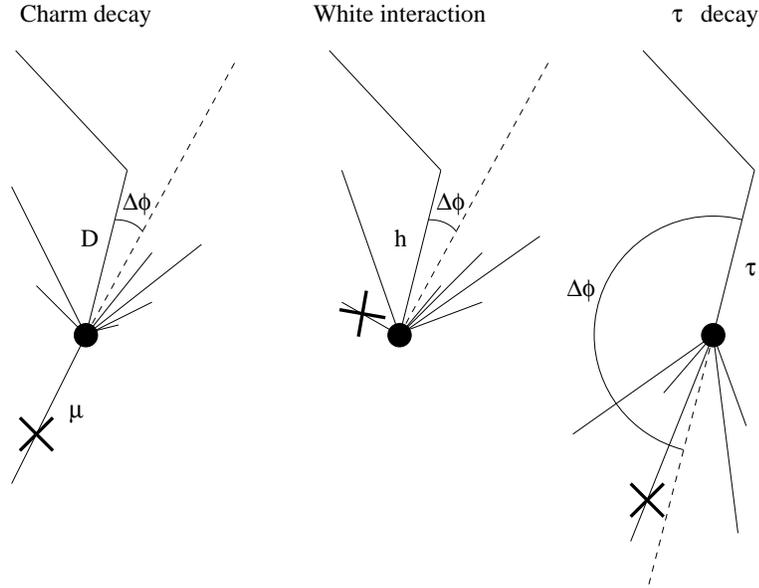,width=10.0cm}}
  \caption{Definition of the angle $\Delta\phi$. The dashed line is the mean direction of the primary tracks, 
with the exclusion of the one which has the largest $\phi$ angle relative to the parent particle. The 
$\Delta\phi$ distribution is different for charm, white interaction or $\tau$ events.}
  \label{fig:Deltaphipicture}
\end{figure}

\begin{figure} [htp]
  \centerline{\psfig{figure=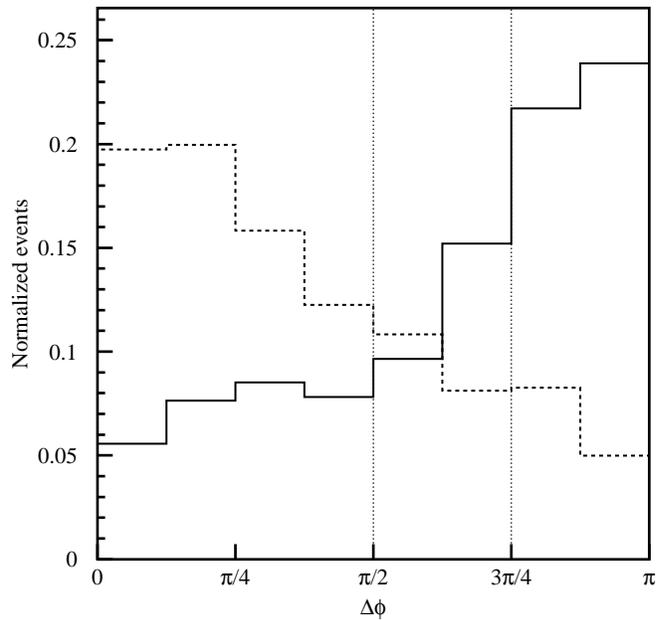,width=10.0cm}}
  \caption{$\Delta \phi$ distributions for simulated $0\mu$ C1 events: comparison between $\nu_\tau$ 
(solid line) and charmed events (dashed line). The area is normalized to 1. 
The vertical lines show the applied cut dividing the events in three subsamples.}
  \label{fig:deltaphicharmtau}
\end{figure}

In addition, the samples have been further subdivided into three $\Delta\phi$ ranges, where $\Delta\phi$ is defined 
as the angle in the transverse plane between the parent track and the mean direction of the other 
primary tracks after exclusion of the one which makes the largest angle with the parent track
(the track with a cross in Fig.~\ref{fig:Deltaphipicture}).
A comparison between the distributions of $\Delta\phi$ in charm and $\tau$ samples is shown in 
Fig.~\ref{fig:deltaphicharmtau}. 
The $\tau$ signal is characterized by a larger mean value of $\Delta\phi$.

\subsubsection{The C3 topology}

\begin{figure} [htp]
  \centerline{\psfig{figure=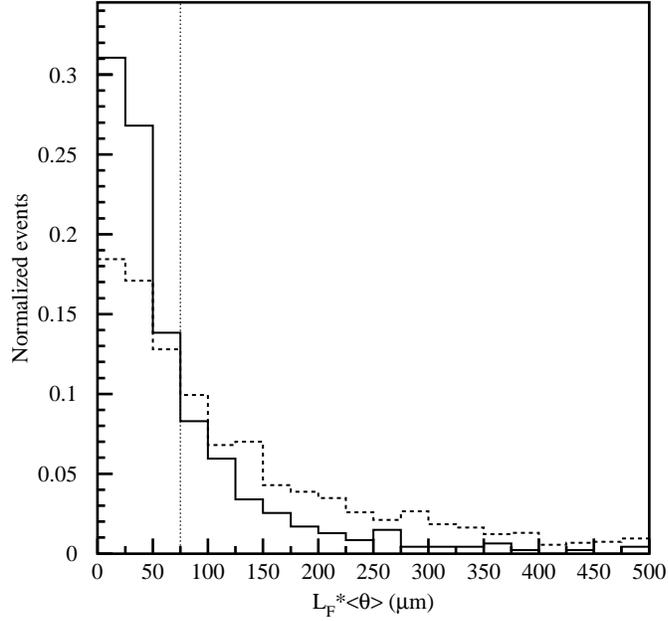,width=10.0cm}}
  \caption{$L_\mathrm{F} \cdot \langle \theta \rangle$ distribution for simulated $0\mu$ C3 events: 
comparison between $\tau$ (solid line) and charm (dashed line) decays. The area is normalized to 1. 
The vertical line shows the applied cut $L_\mathrm{F} \cdot \langle \theta \rangle = 75~\micron$ 
dividing the events in two subsamples.}
  \label{fig:ctau}
\end{figure}

For the C3 topology, the fraction of events for which the three daughter particles have their 
momentum measured is too small to lead to a useful discrimination. Instead, use is made 
of the fact that the mean angle $\langle \theta \rangle$ of the daughter tracks relative 
to the parent is strongly correlated to the inverse of the Lorentz boost factor $\gamma$. 
Thus, the product of the parent flight length $L_\mathrm{F}$ and $\langle \theta \rangle$ 
is a good parameter to discriminate particles of different lifetimes (or apparent 
lifetime in the case of white interactions).
The C3 sample has been divided into ``short $c\tau$'' and ``long $c\tau$'' subsamples, separated 
by a cut value $L_\mathrm{F} \cdot \langle \theta \rangle = 75 \ \micron$. 
Fig.~\ref{fig:ctau} shows a comparison of the charm and $\tau$ samples: the short $c\tau$ sample
is enriched in $\tau$ decays.
These categories are further divided as above according to $\Delta\phi$ ranges. The sample most
sensitive to a $\tau$ signal corresponds to events with short $c\tau$
and large $\Delta\phi$.

Once the different categories are defined the same subdivision is applied 
to the candidate events.
This is done according to the `blind analysis' prescriptions, in order to minimize 
the possibility of biases in the experimental result.

Table~\ref{table:allsubsamples} shows the number of simulated and observed events in C1 
and C3 topologies of the new $0\mu$ sample, together with data from the Phase I analysis
which consist of two samples \cite{osc3}:

\begin{itemize}
\item $1\mu$ events from the 1994--1997 data taking. The background reduction is obtained by applying a cut 
on the transverse momentum ($P_T>250$~\MeVc) of the kink daughter. The kink must occur within five plates 
downstream from the neutrino interaction vertex plate. The estimated background is $0.1$ events and the 
maximum number of $\nu_\tau$ events is $N_\tau^{max} = 5014$ (as reported in Table 2 of \cite{osc3}). The 
background is much smaller than for the $0\mu$ sample because of the low probability of a wrong measurement 
of the charge in the muon spectrometer. No candidate is observed.
\item $0\mu$ one-prong events from the 1994--1995 data taking. Table 2 of \cite{osc3} gives the background and 
the maximum number of $\nu_\tau$ oscillated events for the whole $0\mu$ 1994--1997 sample. The 1996--1997 data 
has been reanalysed, as reported in this paper. The 1994--1995 data correspond to an estimated background of 
$0.3$ events and a maximum number of $\nu_\tau$ oscillated events of $N_\tau^{max} = 526$. No candidate is 
observed.
\end{itemize}

The errors on the expected background are evaluated by taking into account the limited statistics 
of the Monte Carlo sample and the errors on cross-sections and branching ratios.


\section{Results}
\label{sec:results}

From the number of observed candidates, the expected background and the number of signal events 
expected for full oscillation it is possible to compute, for each subsample of 
Table~\ref{table:allsubsamples}, the 90\%~C.L. upper limit on the $\nu_\tau$ 
appearance probability using a frequentist statistical approach, the so-called 
Feldman and Cousins unified approach~\cite{FeldmanAndCousins}.

To evaluate the sensitivity of the experiment, we have simulated a large number of 
experiments taking into account Poisson fluctuations of the expected background. 
The sensitivity is then defined as the average of the 90\%~C.L. upper limits on the 
appearance probability of all these "experiments".

The last two columns in Table~\ref{table:allsubsamples} show the sensitivities 
$S_{\mu\tau}$ of each subsample and an index $i_{\mu\tau}$ in decreasing
order of sensitivity.
The subsamples are then combined, 
with the same prescriptions used by the NOMAD Collaboration \cite{nomad}. 
After including the 11th most sensitive subsample, the global sensitivity changes 
only at the percent level. Similarly for the $\nu_e \rightarrow \nu_\tau$ search. 
In evaluating the global sensitivity, we thus consider only 
the 11 most sensitive subsamples for both searches.


\begin{table}[ht]
  \caption{The final CHORUS data sample. The first two rows refer to the
 Phase I analysis, namely to the $1\mu$ channel 
of the whole data taking (1994--1997) and to the $0\mu$ sample collected in 1994--1995. The new sample, consisting of 
the $0\mu$ data collected in 1996--1997, is divided in C1 and C3 topologies which are further divided in subsamples, 
as described in Section \ref{sec:post-scanning}. 
For each subsample, the following quantities are shown: the expected background; the maximum detectable number of 
$\tau$ events, $N_\tau^{\mu \tau}$ and $N_\tau^{e \tau}$ respectively from the $\nu_\mu$ and $\nu_e$ beam components; 
the number of data events. The numbers in the last two columns are evaluated for the $\nu_\mu \rightarrow \nu_\tau$ 
search (see Section \ref{sec:results}) and give the sensitivity $S_{\mu\tau}$ of each single channel (times $10^4$) 
and an index $i_{\mu\tau}$ which sorts the sensitivities in decreasing order.}
  \begin{displaymath}
    \begin{array}{|c|c||c|c|c|c||r|r|}
      \hline
      \textrm{Category} & \Delta\phi (rad) & \textrm{Background} & N_\tau^{\mu \tau} & N_\tau^{e \tau} & \textrm{Data} & S_{\mu\tau} & i_{\mu\tau} \\
      \hline
      \multicolumn{2}{c}{}\\
      \hline
      \multicolumn{2}{|c||}{\tau \rightarrow 1\mu~[1994-1997~\textrm{data taking}]}             & 0.100\pm0.025&  5014 &  55.8   &   0  & 4.9 & 1 \\
      \hline
      \multicolumn{2}{|c||}{\tau \rightarrow 0\mu~\textrm{C1}~[1994-1995~\textrm{data taking}]} & 0.300\pm0.075&   526 &  5.85   &   0  & 48  & 10 \\
      \hline
      \multicolumn{2}{c}{}\\
      \hline
      \multicolumn{2}{|c||}{\tau \rightarrow 0\mu~\textrm{C1}~[1996-1997~\textrm{data taking}]} & 53.2\pm9.0   &  9621 &  76.9 & 59 &   & \\
      \hline
                                                        & [0;\pi/2]          & 18.0\pm2.3      &   769 & 7.67 & 26 & 211& 15 \\

      \textrm{No momentum measured}                     & [\pi/2;3\pi/4]     &  5.00\pm0.73    &   708 & 5.34 & 10 & 66 & 12 \\

                                                        & [3\pi/4;\pi]       &  6.2\pm1.4      &  1406 & 13.9 &  7 & 38 & 6 \\
      \hline
      \textrm{Only MCS momentum measured:}              & [0;\pi/2]          &  4.6\pm1.1      &   991 & 6.75 &  2 & 45 & 9 \\

      P^{MCS}_T>250~\MeVc                               & [\pi/2;3\pi/4]     &  1.20\pm0.40    &   749 & 5.70 &  2 & 34 & 5 \\

      \textrm{and ($P^{MCS}$ vs. $L_\mathrm{F}$) cut}   & [3\pi/4;\pi]       &  3.3\pm1.0      &  1649 & 12.8 &  3 & 26 & 4 \\
      \hline
      \textrm{DT momentum measured (Charge -):}         & [0;\pi/2]          &  0.383\pm0.071  &   546 & 3.62 &  0 & 41 & 8 \\

      P^{DT}_T>250~\MeVc                                & [\pi/2;3\pi/4]     &  0.087\pm0.033  &   556 & 4.24 &  0 & 38 & 7 \\

      \textrm{and ($P^{DT}$ vs. $L_\mathrm{F}$) cut}    & [3\pi/4;\pi]       &  0.055\pm0.012  &  1023 & 7.24 &  0 & 22 & 2 \\
      \hline
   \multicolumn{2}{|c||}{P_T<250~\MeVc\textrm{ or Charge + or in region II}} &  14.6\pm1.6     &  1224 & 9.77 &  9 &    & \\

      \hline
      \multicolumn{2}{c}{}\\
      \hline

      \multicolumn{2}{|c||}{\tau \rightarrow 0\mu~\textrm{C3}~[1996-1997~\textrm{data taking}]} &  47\pm11     &  4443 & 35.5 &  48 & & \\
      \hline
                                       & [0;\pi/2]          &  14.8\pm5.0     &   792 & 6.63 & 17 & 133 & 14 \\

      \textrm{Short}~c\tau~(<75~\micron)  & [\pi/2;3\pi/4]     &   6.4\pm3.0     &   782 & 6.12 &  6 &  89 & 13 \\

                                       & [3\pi/4;\pi]       &   1.5\pm1.5     &  1554 & 13.4 &  4 &  23 &  3 \\
      \hline
                                       & [0;\pi/2]          &  15.5\pm5.0     &   386 & 2.37 &  8 & 268 & 17 \\

      \textrm{Long}~c\tau~(>75~\micron)   & [\pi/2;3\pi/4]     &   9.8\pm3.9     &   336 & 2.40 &  8 & 237 & 16 \\

                                       & [3\pi/4;\pi]       &   1.7\pm1.5     &   593 & 4.62 &  5 &  62 & 11 \\
      \hline
    \end{array}
  \end{displaymath}
  \label{table:allsubsamples}
\end{table}


\subsection{$\nu_\mu \rightarrow \nu_\tau$ oscillation}

Fig.~\ref{fig:sensy} (left) shows the distribution of the upper limits (at 90\%~C.L.) obtained 
by simulating 800 experiments. The average value, $S = 2.4 \times 10^{-4}$, corresponds 
to the CHORUS sensitivity to $\nu_\tau$ appearance.

The 90\%~C.L.\ upper limit on the appearance probability obtained from
the data is

$$P (\nu_\mu \rightarrow \nu_\tau) < 2.2 \times 10^{-4} \ ,$$

\noindent indicated by a vertical line in Fig.~\ref{fig:sensy}. It is in agreement with the expected 
sensitivity. In fact, in absence of signal events the probability to obtain an upper limit 
of $2.2 \times 10^{-4}$ or lower is $48\%$, which means that the number of observed events 
is compatible with the estimated background. The above result improves by a factor $1.5$ 
the previously published limit $P<3.4\times 10^{-4}$ \cite{osc3}.

In a two-neutrino formalism, the above upper limit corresponds 
to the exclusion region in the ($\Delta m^2$, $\sin^2 2\theta_{\mu\tau}$) oscillation parameter
plane shown in Fig.\ref{fig:exclusionplot} (left) and to the limit
$\sin^2 2\theta_{\mu\tau} < 4.4 \times 10^{-4}$ for large $\Delta m^2$.

The upper limit obtained by NOMAD \cite{nomad} is more stringent than the CHORUS one at
large $\Delta m^2$, whereas for $\Delta m^2$ lower than $70$~\eVeVcccc the upper limit is 
improved by CHORUS owing to its higher efficiencies at low neutrino energies.


\subsection{$\nu_e \rightarrow \nu_\tau$ oscillation}

The SPS neutrino beam contains a 0.8\% $\nu_{e}$ contamination. 
Assuming that all observable $\nu_\tau$ would originate from this contamination, 
the above result translates into a limit on the $\nu_{e} \rightarrow \nu_\tau$ appearance 
probability. The difference in energy between the $\numu$ ($\langle E_{\numu} \rangle \sim 26$ \GeV) 
and $\nu_e$ ($\langle E_{\nu_e} \rangle \sim 42$ \GeV) components leads 
to a different shape of the exclusion plot in the oscillation parameter plane.

\begin{figure}[htp]
  \begin{center}
    \rotatebox{0}{
      \resizebox{1.0\textwidth}{!}{ 
	\includegraphics{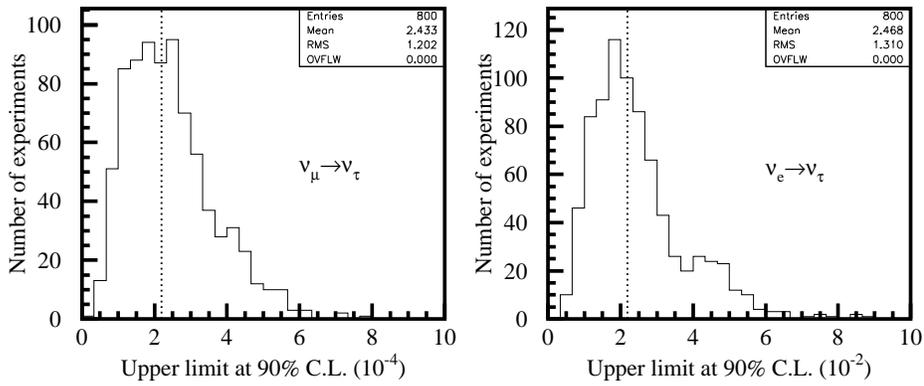}    } }
    \caption{Upper limits obtained at 90\%~C.L. \cite{FeldmanAndCousins}, in absence of signal events, 
for 800 simulated experiments with the CHORUS expected background. The average value corresponds to the 
sensitivity to $\nu_\mu \rightarrow \nu_\tau$ (left) and $\nu_e \rightarrow \nu_\tau$ (right) appearance
probability. The vertical line is the 90\%~C.L. upper limit obtained from the data.}
    \label{fig:sensy}
  \end{center}
\end{figure}

Fig.~\ref{fig:sensy} (right) shows the upper limit distribution obtained by simulating 800 experiments.
The average value, $S = 2.5 \times 10^{-2}$, corresponds to the
sensitivity to the $\nu_e \rightarrow \nu_\tau$ appearance probability. 

\begin{figure}[htp]
  \begin{center}
      \resizebox{1.0\textwidth}{!}{
	\includegraphics{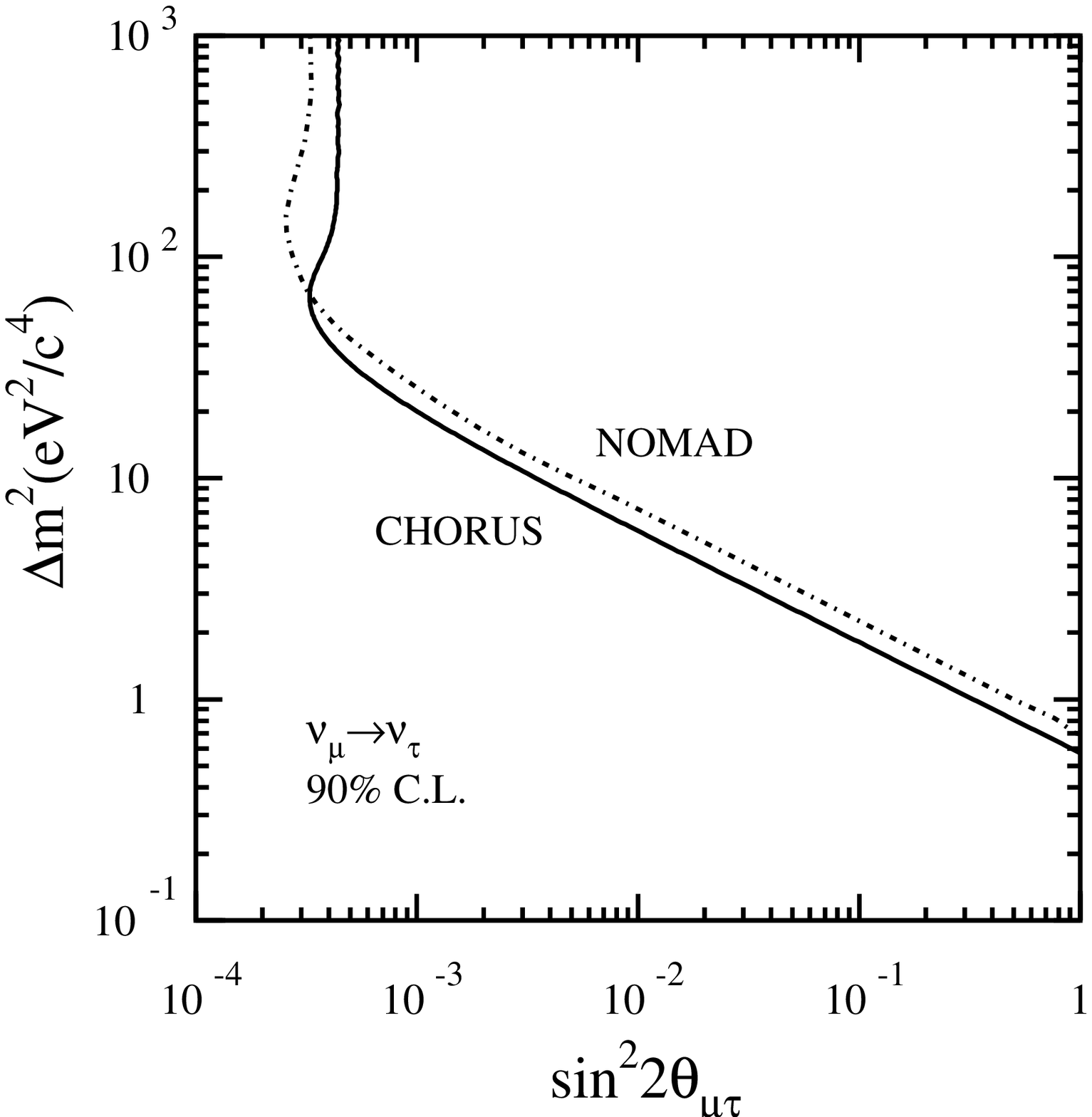} \includegraphics{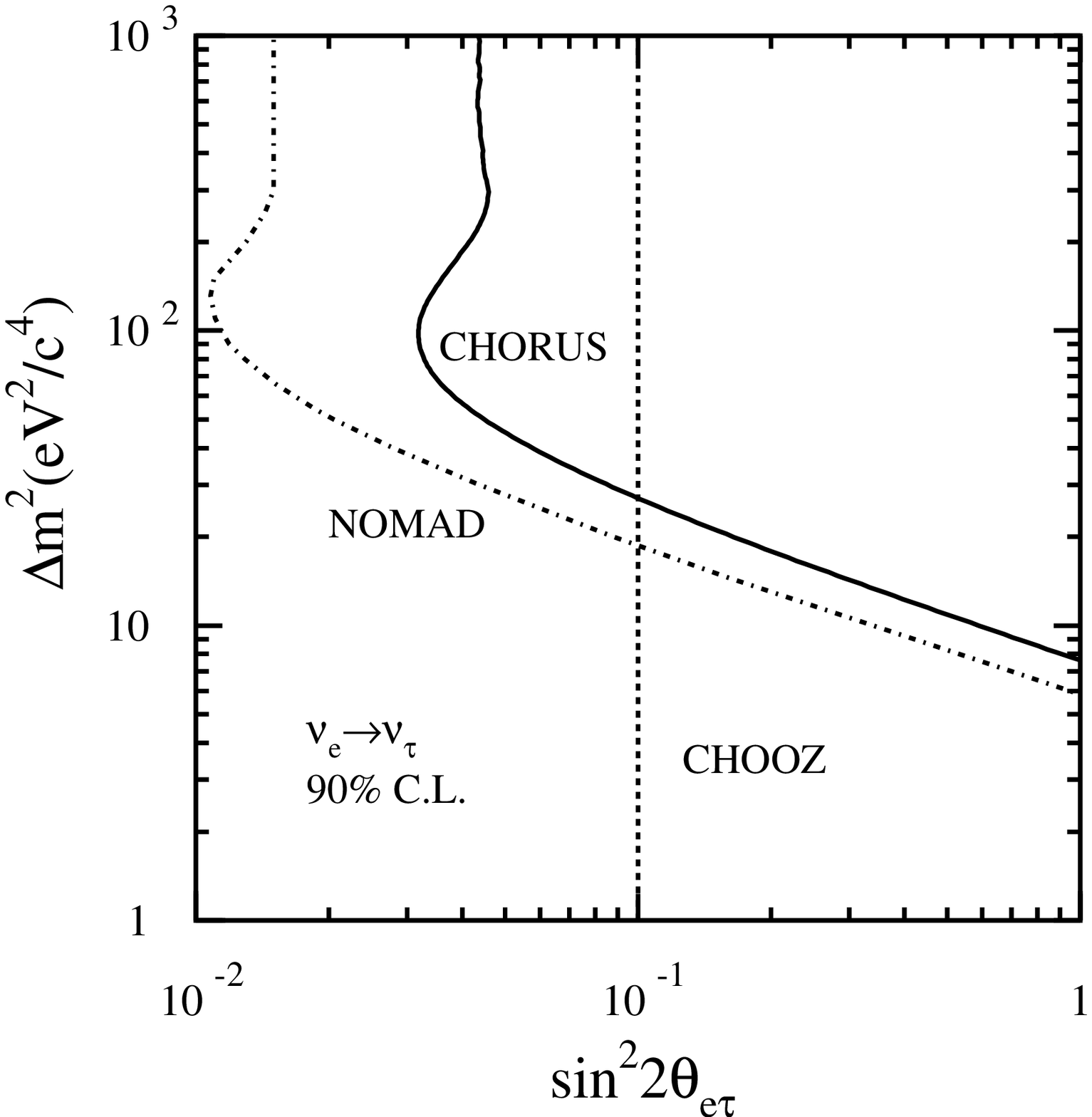}    }
    \caption{The CHORUS upper limit on $\nu_\mu \rightarrow \nu_\tau$ (left) and $\nu_e \rightarrow \nu_\tau$ 
(right) oscillation represented in an exclusion plot in the oscillation parameter plane. CHORUS results are 
shown as solid lines and are compared with the last results of NOMAD \cite{nomad} and CHOOZ \cite{CHOOZ}. }
    \label{fig:exclusionplot}
\end{center}
\end{figure}

The 90\%~C.L.\ upper limit on the appearance probability obtained from the data is

$$P (\nu_e \rightarrow \nu_\tau) < 2.2 \times 10^{-2} \ ,$$

\noindent indicated by a vertical line in Fig.~\ref{fig:sensy}.
In absence of signal events, the probability to obtain an upper limit of $2.2 \times 10^{-2}$ 
or lower is $49\%$. 
The above result improves by a factor $1.2$ the previously published limit~\cite{osc3}.
The improvement is smaller than in the $\nu_\mu \rightarrow \nu_\tau$ case for two reasons.
The first one is the variation of the expected $\nu_{e}$ 
contamination provided by a new simulation of the beam (reduced by a factor 0.85) \cite{gbeam}.
Furthermore, in the  $\nu_\mu \rightarrow \nu_\tau$ case the efficiency improved predominantly 
at low neutrino energies and the harder $\nu_e$ spectrum makes this improvement less effective. 
This reason also explains the large difference in the limit relative to NOMAD.
In a two-neutrino formalism, the above upper limit corresponds to the exclusion region 
in the ($\Delta m^2$, $\sin^2 2\theta_{\mathrm{e}\tau}$) oscillation parameter plane shown 
in Fig.\ref{fig:exclusionplot} (right) and to the limit 
$\sin^2 2\theta_{e\tau} < 4.4 \times 10^{-2}$ for large $\Delta m^2$.


\section{Conclusions}
The final result of the search for $\nu_\tau$ appearance with the CHORUS experiment has been 
presented. 
A 90\%~C.L.\ upper limit of $2.2 \times 10^{-4}$ was set for the appearance probability, 
improving by a factor 1.5 the previously published CHORUS result. 
In a two-neutrino mixing scheme, this result corresponds 
to the limit $\sin^2 2\theta_{\mu\tau} < 4.4 \times 10^{-4}$ for large $\Delta m^2$. 
With respect to previous CHORUS results, we have also improved by a factor 1.2 the upper limit 
on the ($\nu_e \rightarrow \nu_\tau$) oscillation probability.


\section*{Acknowledgements}
We gratefully acknowledge the help and support of the neutrino beam
staff and of the numerous technical collaborators who contributed to
the detector construction, operation, emulsion pouring, development,
and scanning.  The experiment has been made possible by grants from
the Institut Interuniversitaire des Sciences Nucl\'eair\-es and the
Interuniversitair Instituut voor Kernwetenschappen (Belgium), the
Israel Science Foundation (grant 328/94) and the Technion Vice
President Fund for the Promotion of Research (Israel), CERN (Geneva,
Switzerland), the German Bundesministerium f\"ur Bildung und Forschung
(contract numbers 05 6BU11P and 05 7MS12P)
(Germany), the Institute of Theoretical and Experimental Physics
(Moscow, Russia), the Istituto Na\-zio\-na\-le di Fisica Nucleare
(Italy), the Promotion and Mutual Aid Corporation for Private Schools
of Japan and Japan Society for the Promotion of Science (Japan), the
Korea Research Foundation Grant (KRF-2003-005-C00014) (Republic of
Korea), the Foundation for Fundamental Research on Matter FOM and the
National Scientific Research Organization NWO (The Neth\-er\-lands),
and the Scientific and Technical Research Council of Turkey
(Turkey). We gratefully acknowledge their support.

\end{document}